# Structures of cycloserine and 2-oxazolidinone probed by X-ray photoelectron spectroscopy

Marawan Ahmed[1], Feng Wang[1], Robert G. Acres[2], Kevin C. Prince[2,3,1,$]

[1] eChemistry Laboratory, Faculty of Life and Social Sciences, Swinburne University of Technology, Melbourne, Victoria 3122, Australia,
[2] Elettra-Sincrotrone Trieste, in Area Science Park, I-34149 Basovizza, Trieste, Italy,
[3] Istituto Officina dei Materiali, Consiglio Nazionale delle Ricerche, Area Science Park, I-34149 Trieste, Italy.
[$] To whom correspondence should be addressed. Phone: +39 0403758584. Fax: +39 0403758565. E-mail: Kevin.Prince@elettra.trieste.it.

Supporting Information Placeholder

**Abstract**

The electronic structures and properties of 2-oxazolidinone and the related compound cycloserine (CS) have been investigated using core and valence photoelectron spectroscopy and theoretical calculations. Isomerization of the central oxazolidine heterocycle and the addition of an amino group yields cycloserine. Theory correctly predicts the C, N and O 1s core spectra, and additionally we report theoretical natural bond orbital (NBO) charges. The valence ionization energies are also in agreement with theory and previous measurements. Although the lowest binding energy part of the spectra of the two compounds show superficial similarities, analysis of the charge densities of the frontier orbitals indicates substantial reorganization of the wave functions as a result of isomerization. The Highest Occupied Molecular Orbital of CS has leading carbonyl π character with contributions from other heavy atoms in the molecule, while the Highest Occupied Molecular Orbital of 2-oxazolidinone has leading nitrogen, carbon and oxygen pπ character. At laboratory temperatures, CS is predicted to exist mainly as two low energy conformers in the gas phase, separated in energy by 4.6 kJ•mol$^{-1}$, whereas 2-oxazolidinone is present as a single conformer in the gas phase. However the experimental spectra do not provide proof of the presence of these conformers. Theoretical predictions of resonance effects are supported by the photoelectron structure and by published crystallographic data, yielding a model that is consistent with the valence spectra, geometric structure and core level spectra.

# 1. INTRODUCTION

In this paper, we investigate the electronic structure of two heterocyclic organic compounds of pharmacological interest, cycloserine (systematic name (4R)-4-amino-1,2-oxazolidin-3-one), denoted CS here, and 2-oxazolidinone (systematic name 1,3-oxazolidin-2-one), or OX2, Figure 1. These compounds are closely related chemically as their central rings are isomeric, and both have an oxo side group, while cycloserine has an additional amino side group.

Cycloserine was introduced as an antibiotic for treatment of tuberculosis and is still used as a "second line" drug for this purpose.[1,2] Its use is restricted because of its toxic and psychotropic side effects, but it has found applications in treatment of central nervous system disorders. The other compound examined here, OX2, is also a heterocycle and its derivatives are used as drugs for treatment of MRSA (Methicillin-Resistant Staphylococcus Aureus), one of the greatest problems in modern hospitals.[3-5] Other drugs are under development, based on this core chemical group.[6,7] The present work also builds on our recent study of some other drug-related, heterocyclic compounds, including cyclopeptides, thiazolidine carboxylic acids and 2-azetidinone.[8-10]

The crystal structure of the hydrochloride salt of CS has been determined by Turley et al.[11] It has been shown using IR spectroscopy that CS can be dimerized in the solid state, and also that its form in solution is mostly pH dependent.[12,13] In an interesting DFT (density functional theory) study, Yosa et al. showed that the biological activities of some NMDA (*N*-methyl-D-aspartate) receptor ligands, including CS, can be determined by investigating the HOMOs (Highest Occupied Molecular Orbitals) and the LUMOs (Lowest Unoccupied Molecular Orbitals) of the compounds.[14] This represents the fulfillment of a goal of an early valence band photoelectron study of OX2,[15] which determined ionization potentials with a view to later correlating the data with medicinal activity. In the solid state the molecular structure is constrained, but in the gas phase greater conformational freedom is available. By analogy with amino acids, the primary amine group in CS may be expected to adopt various conformations, and we have investigated this question.

Our second compound, OX2, has been well-studied,[16, 17] and OX2 and its derivatives are widely used in the pharmaceutical industry.[17] It is a building block of the active antibacterial agent Linezolid[18], and its derivatives have powerful catalytic activity in Aldol-type reactions.[19, 20] The X-ray crystal structure of OX2 shows evidence of the presence of intermolecular hydrogen bonding.[21, 22] The valence spectrum of OX2 has been reported previously,[15, 23] but we are not aware of any such studies of CS, and there appear to be no gas phase core ionization studies for either OX2 or CS. Andreocci et al.[24] measured the condensed phase core spectra of OX2 and a number of related compounds, and discussed the effect of resonance structures and delocalization of charge, a subject to which we will return later.

There are few computational studies of these important molecules. The current study focuses on investigating the chemical structures of these two molecules in the gas phase using synchrotron sourced soft X-ray photoelectron spectroscopy for the core and valence shells, together with theoretical calculations.

## 2. EXPERIMENTAL AND COMPUTATIONAL DETAILS

The measurements were performed at the Gas Phase photoemission beamline of Elettra, Trieste, Italy, using apparatus and calibration methods described previously.[25-28] The total resolution of the photons and analyzer was estimated to be 0.2, 0.32, 0.46, and 0.78 eV at photon energies of 100 (valence band), 382 (C 1s), 495 (N 1s), and 628 eV (O 1s), respectively. The compounds were supplied by Sigma Aldrich (purity > 98%) and used without further purification. The D enantiomer of cycloserine was used, but since the measurements and calculations were not sensitive to chiral properties, the compound is described in the following as simply cycloserine. The samples were evaporated from a non-inductively wound furnace at temperatures of 395 K for CS and 320 K for OX2. They were checked for evidence of thermal decomposition, such as spectral changes as a function of time, presence of decomposition products which are easily identified in valence spectra ($H_2O$, $CO_2$, etc), discoloration after heating, etc. No evidence was found for decomposition of the compounds.

All geometry optimizations and calculations of natural bond orbital (NBO)[29] charges were performed using the density functional theory (DFT) based B3LYP/6-311++G** model, which is incorporated in the Gaussian 09 (G09)[30] computational chemistry package. We examined the conformers which can be generated by rotating the $N_{(7)}$ amino group. Three low energy conformers of CS, denoted CS-**I**, CS-**II** and CS-**III** and shown in Figure 2, were identified on the potential energy surface (PES) using a relaxed PES scan around the $C_{(3)}$- $C_{(4)}$-$N_{(7)}$-H dihedral angle at the B3LYP/6-31G* level of theory. The B3LYP model has proven to provide reliable geometries for several bio-molecules.[31-33] The outer valence Green's function (OVGF)[34, 35] theory, which is coupled with the 6-311++G** basis set, was used to calculate valence spectra. The OVGF model, which is incorporated in the G09 computational chemistry package, accurately predicts outer-valence ionization potentials.[10, 36-39]

The calculations of the binding energy (ionization potential, IP) spectra were carried out using the Amsterdam Density Functional (ADF) computational chemistry program[40-42] except for the OVGF outer-valence calculations. Core orbital calculations were performed by applying the $\Delta E_{KS}$ method and the conventional LB94 model. The $\Delta E_{KS}$ method determines the difference in the total Kohn-Sham energies between the core-ionized cation and the neutral parent molecule using the (PW86-PW91)/et-pVQZ model,[43] thereby taking account of the core hole relaxation effects. The vertical core ionization potentials of OX2 and CS were calculated using the LB94/et-pVQZ[44] model for the core shell, and the vertical valence ionization spectra were calculated using the SAOP/et-pVQZ model.[45] Here the basis set et-pVQZ is an even-tempered polarized-valence quadruple-zeta Slater type basis set.[46] The "meta-Koopmans' theorem"[47] was applied without any further modifications and scaling. Several other software packages including Gaussview 5.0,[48] ADFVIEW[40-42] and Molden[49] were employed in the present study for visualization.

## 3. RESULTS AND DISCUSSION

### 3.1. Geometry and NBO charges

The results of the structural calculations are compared in Table 1 with published theoretical and x-ray crystallographic data.[13, 17, 18] The perimeter of the ring is correctly described within 0.5 to 1.5% for the two compounds, however bond angles show certain discrepancies from the reported crystal structures as a result of the absence of intermolecular and crystal packing forces in the gas phase. Apart from this discrepancy, the overall agreement is good. A three dimensional rendering of the structures is shown in Figure 2, with NBO charges on each atom shown in parentheses. The PES scan of CS is given in the supplementary materials (S1).

The minimum energy conformer was found to be CS-**I**. The conformers CS-**II** and CS-**III** have Gibbs free energies (ΔG) which are respectively 4.3 and 11.6 kJ·mol$^{-1}$ higher than the ground state at the experimental temperature of 395 K. This yields Boltzmann populations of 77, 21 and 2%. The barriers for conversion, **ΔTE+ZPE**, between the conformers are about 4.6 and 12 kJ·mol$^{-1}$, which implies facile conversion at 395 K. The relative energies are given in Table S3 of the Supplementary Materials.

In CS-**I**, the $N_{(7)}$ amino group is oriented toward the ring, forming two intramolecular hydrogen bonds (H-bonds), namely $N_{(7)}$-H···$O_{(6)}$ (2.966 Å) and the $N_{(7)}$-H'···$O_{(1)}$ (2.915 Å). On the other hand, for CS-**II** and CS-**III**, the –NH$_2$ group is oriented towards the other side forming the $N_{(7)}$-H···$O_{(6)}$ bond in CS-**II** of length 2.704 Å, and a $N_{(7)}$-H'···$O_{(1)}$ bond in CS-**III** of length 2.963 Å. Also, the degree of puckering, estimated by the puckering amplitude ($v_{max}$),[50] is higher in CS-**III** (40.16) than CS-**I** (37.96) and CS-**II** (34.49). Conformer **I** has the lowest free energy, implying that the energy gain due to the formation of two slightly longer hydrogen bonds outweighs the advantage of a shorter but single bond.

Since $C_{(3)}$ is sp$^2$ hybridised $N_{(2)}$, $C_{(3)}$, $O_{(6)}$ and $C_{(4)}$ are coplanar (the CS-**I** $N_{(2)}$-$C_{(3)}$-$O_{(6)}$-$C_{(4)}$ dihedral angle is -179.4°), with a $C_{(4)}$-$C_{(3)}$-$N_{(2)}$ angle of approximately 120°. Since these three atoms are members of a five rather than six membered ring, the angle is in fact smaller, 112.2° (experiment) or 106.13° (theory, CS-**I**). This geometry favors conjugation with the lone pair orbital of $N_{(2)}$ and the formation of partial $C_{(3)}$-$N_{(2)}$ double bond character. The calculated bond length has a value of 1.38 Å, compared with an experimental value of 1.299 Å, see Table 1. Both the angle and the bond length indicate that theory underestimates slightly the degree of conjugation. However, the discrepancy between theory and experiment, can again be due to the crystal phase of the experimental sample.

The geometry of OX2 is even more favorable for conjugation, with the carbonyl $C_{(2)}$ sp$^2$ hybridized, and bonded to $O_{(1)}$ and $N_{(3)}$, ($O_{(1)}$-$C_{(2)}$-$O_{(6)}$-$N_{(3)}$ dihedral angle -179.41°) giving the possibility of hyperconjugation. The ring perimeter ($R_5$) of OX2 is shorter than the values for CS, theoretically by 0.138 Å and experimentally by 0.145 Å. This may be due to the existence of two resonating groups in OX2, the peptide-like ($N_{(3)}C_{(2)}$=$O_{(6)}$) group and the ester-like ($O_{(1)}C_{(2)}$=$O_{(6)}$) group, which make OX2 a dual entity being a lactam (cyclic amide) and a lactone (cyclic ester). This may also explain the lower theoretical puckering amplitude $v_{max}$ of OX2 in the gas phase, 26.49, compared with 37.96 (CS-**I**), 34.49 (CS-**II**) and 40.16(CS-**III**), as resonance may impose a certain degree of planarity. However, due to the inherent flexibility of these molecules and the absence of any crystal packing forces in gas phase calculations,[51] it is observed that the calculated puckering amplitudes are slightly higher than the experimentally derived values for CS conformers.

For OX2, previous studies emphasized the presence of a $N_{(3)}$-H···$O_{(6)}$ intermolecular H-bond[21, 22] but this is not relevant to the present case. Although the intramolecular $N_{(3)}$-H···$O_{(6)}$ distance is about 2.599 Å, the presence of an intramolecular H-bond is unlikely because of the sharp angle along this axis (63.02°).

Figure 2 also presents the atomic site based NBO charges of the molecules at the B3LYP/6-311++G** level of theory. The atomic charge distribution in both molecules is consistent with basic electronegativity arguments. All the non-carbon heavy atoms (N's and O's) possess negative NBO charges. For the CS conformers, the orientation of the amino group -NH$_2$ does not affect the NBO charges of the atoms on the backbone frame significantly. However, the NBO charges on the $O_{(6)}$ in CS-**I** and CS-**II** are closer to each other than to conformer CS-**III**, whereas the NBO charges on the $O_{(1)}$ and $N_{(7)}$ sites of CS-**I** and CS-**III** are more similar to those of conformer CS-**II**. For example, the NBO charges on the $O_{(6)}$ atom of CS-**I**, CS-**II** and CS-**III** are -0.597 *e*, -0.596 *e* and -0.572 *e*, respectively. When inside of the pentagon ring, the $O_{(1)}$ atom possesses NBO charges of -0.428 *e*, -0.418 *e* and -0.429 *e*, respectively, in CS-**I**, CS-**II** and CS-**III**. The NBO charges of the $N_{(7)}$ atoms of CS-**I**, CS-**II** and CS-**III** are -0.826 *e*, -0.843 *e* and -0.827 *e*, respectively. This is because the three dimensional orientation of the amino group $-NH_2$ in the conformers permits different H-bonds. Conformers CS-**II** and CS-**III** permit only one intramolecular H-bond, i.e., $N_{(7)}$-H···$O_{(6)}$ for CS-**II** (2.704 Å) and $N_{(7)}$-H'···$O_{(1)}$ for CS-**III** (2.963 Å). In CS-**I** conformer, however, both H-bonds shown in its structure: an $N_{(7)}$-H···$O_{(6)}$ (2.966 Å) connects with the carbonyl oxygen, and the other H-bond at $N_{(7)}$-H'···$O_{(1)}$ (2.519 Å) is associated with the oxygen in the pentagon ring. As a result, the NBO charges reflect the H-bond(s) of the CS conformers.

The carbon atoms in the CS conformers present a different picture as the carbon atoms may be either positively charged, e.g., the carbonyl carbon $C_{(3)}$ with 0.642 *e* (CS-**I**), 0.663 *e* (CS-**II**) and 0.657 *e* (CS-**III**), or negatively charged, e.g., $C_{(4)}$ and $C_{(5)}$, depending on the NBO charges of their neighboring atoms. The carbonyl carbon $C_{(3)}$ is positively charged in all CS conformers as it directly bonds with two negatively charged atoms, i.e., $O_{(6)}$ and $N_{(2)}$. However, $C_{(4)}$ and $C_{(5)}$ only directly bond with one negatively charged atom, either $N_{(7)}$ or $O_{(1)}$. As a result, the NBO charges suggest that the C 1s spectrum of CS can consist of two peaks: one due to $C_{(3)}$, and the other a broader peak due to $C_{(5)}$ and $C_{(4)}$.

For OX2, as stated above, the amino group bonded to $C_{(4)}$ in CS is replaced by a hydrogen atom, and the carbonyl carbon and the nitrogen atom switch positions in the pentagonal ring (isomerization). It is clear that no intramolecular H-bond can form and the calculated structure of OX2 confirms this. The carbonyl carbon, $C_{(2)}$, has a significantly higher NBO charge of 0.912 *e* than the counterpart of CS carbonyl carbon, $C_{(3)}$ (0.54 to 0.66) e, due to the fact that $C_{(2)}$ is bound to two oxygen atoms. The NBO charges of the $C_{(4)}$ and $C_{(5)}$ sites in OX2 are -0.201 *e* and -0.031 *e*, respectively. The NBO charges also suggest two spectral peaks in the C 1s spectra, one for $C_{(2)}$ and a broader peak for $C_{(4)}$ and $C_{(5)}$.

**3.2. Core ionization spectra**

Table 2 lists the measured and calculated core ionization energies of the molecules at different levels of theory. In the table, the IPs of the two most stable CS conformers (CS-**I** and CS-**II**) are listed, but discussion will be limited to the more stable conformer CS-**I**. Figure 3 (upper panel) compares the measured C 1s ionization spectra of CS-**I** with the spectrum calculated using the $\Delta E_{ks}$ method, and they agree well. The single high energy peak is assigned to the carbonyl carbon $C_{(3)}$, whereas the lower energy peak is assigned to the $C_{(5)}$ and $C_{(4)}$ atoms with a theoretical splitting of 0.53 eV. This low energy feature is structured, implying two components as expected. The order of core ionization energies in CS is $C_{(3)}>C_{(5)}>C_{(4)}$, which is predicted by the NBO charges. Theoretical IP values have the same ordering from both the LB94 and the $\Delta E_{ks}$ models.

Figure 3 (lower panel) presents the C 1s core ionization spectra of OX2. The order of the core level ionization potentials is $C_{(2)}>C_{(5)}>C_{(4)}$, which is consistent with the NBO charges. This indicates that in OX2, core ionization is dominated by initial state effects. The $C_{(4)}$ and $C_{(5)}$ binding energies are similar to those in CS, consistent with the fact that the chemical environment of these two carbons is rather similar to that in CS. The theoretical energy splitting is 0.33 eV, which is a little lower than the energy splitting between the two lower binding energy carbons in the spectrum of CS-**I** ($C_{(4)}$ and $C_{(5)}$), 0.53 eV. The experimental core IP of the carbonyl carbon in OX2 is higher than that in CS-**I** by about 1.6 eV (theory, 1.3 eV), because this carbon is directly bonded to three electronegative atoms ($O_{(1)}$, $N_{(3)}$ and $O_{(6)}$), whereas in CS, it is directly bonded to only two electronegative atoms ($N_{(2)}$, and $O_{(6)}$).

Figure 4 displays the N 1s core ionization spectra of CS-**I** (upper panel) and OX2 (lower panel). For CS, the peak at higher energy, IP = 407.2 eV, is assigned to the ring secondary amino nitrogen ($N_{(2)}$), while the peak at 405.6 eV is assigned to the primary amino nitrogen ($N_{(7)}$). The lower panel displays the N 1s spectra of OX2. A notable feature of the spectra is that in OX2, the N 1s IP has a value between the two N 1s core IPs of CS-**I** (406.2 eV), and this is directly related to the local chemical environment of each nitrogen atom. In CS, the ring nitrogen $N_{(2)}$ is bonded to a carbonyl group ($C_{(3)}=O_{(6)}$) and an oxygen atom $O_{(1)}$, while the amino nitrogen $N_{(7)}$ is not directly bonded to electronegative groups/atoms. In OX2, $N_{(3)}$ is bonded to one carbon atom and the carbonyl group ($C_{(2)}=O_{(6)}$). Thus we expect that the binding energy of $N_{(3)}$ in OX2 is less than the binding energy of $N_{(2)}$ in CS-**I**. However the question arises as to why the binding energy of $N_{(3)}$ in OX2 is higher than that of $N_{(7)}$ in CS-**I** since both are amino nitrogen atoms bonded to carbon atoms. We attribute the difference to resonance interaction in OX2 between $N_{(3)}$ and $O_{(6)}$, and possibly also $O_{(1)}$. In CS-**I** the nitrogen is a third nearest neighbor of oxygen, whereas in OX2 the two atoms are second nearest neighbours, the arrangement required for maximizing resonance interaction.[52]

Figure 5 presents the measured O 1s spectra of CS-**I** (upper panel) and OX2 (lower panel). Theoretical spectra are shifted by +0.5 eV (CS-**I**) and +0.3 eV (OX2) to match the experiment. Experimentally, the higher energy oxygen peaks, IP = 539.85 eV (CS-**I**), 539.46 eV(OX2) are assigned to the ring oxygen $O_{(1)}$, and the lower energy peaks at 537.75 eV (CS-**I**) and 537.55 eV (OX2), are assigned to the carbonyl oxygen $O_{(6)}$.

Compared to OX2, the $O_{(1)}$ IP of CS-**I** is higher by about 0.39 eV ($\Delta E_{ks}$ and experiment), which may be explained by the existence of partial double bond character in OX2 (the $O_{(1)}$-$C_{(2)}$ distance is 1.371 Å) due to resonance with the carbonyl group, that is directly attached to this oxygen in OX2 but not in CS.

### 3.3. Valence ionization spectra

The calculated and measured valence band energies of both compounds are displayed in Table 3, and the data are plotted in Figures 6 and 7. For CS, the theoretical curve is the Boltzmann weighted sum of the two lowest energy conformers, CS-**I** and CS-**II**. The spectroscopic pole strengths calculated using the OVGF model are all between 0.89 and 0.91, indicating that the single particle approximation used in the models is valid.

The features of the valence spectrum of cycloserine (Figure 6) are labelled from A to N. A number of weak features appear due to the residual gas in the experimental chamber, which consists mostly of water (feature d') and nitrogen (h', i'). These are easily identified by their very narrow line widths, typical of small molecules. In Table 3, the vertical ionization energies are compared with values calculated using the SAOP and OVGF methods. The first seven valence states observed experimentally can be unambiguously assigned to ionization of molecular orbitals 27a to 21a, as calculated by OVGF. No clear features appear which can be associated with orbitals 20a, 19a, 17a and 16a, but in the experimental spectrum, there is clearly intensity in the regions where these ionic states are expected. We conclude that these particular states are either broadened so much that they cannot be distinguished, or that they overlap two-hole one-particle states which obscure them, or that both effects are present. Valence states 18a and 15a are clearly identified as well as those mentioned above.

The theoretical spectrum takes into account two conformers weighted according to the calculated populations, but cross-sections have not been calculated. Thus quantitative agreement with the intensities is not expected, and we focus on the energies of the features. There is no unambiguous evidence for the presence of two conformers, and the spectrum can be explained assuming a single conformer CS-**I**.

For the inner valence region at higher binding energies, the Outer Valence Green's Function calculations do not make predictions and we turn to the SAOP calculations. This method allows us to assign the single-hole states in this region, although there are also many two-hole, one-particle states as well, which tend to add a continuum background to the spectrum. The features J to N can be assigned to valence singly ionized states. No clear features were identified which can be associated with the valence orbitals 11a and 9a, and again we ascribe this to effects of broadening or overlap with excited states.

Considering the first 9 predicted and observed outer valence ionic states, the OVGF formalism predicts the energies with a mean deviation from experiment of 0.22 eV. The SAOP calculations fare worse, with a mean deviation of 0.49 eV from the experimental values, and the main discrepancies here are for the HOMO and HOMO-3 molecular orbitals. At higher binding energies, the agreement improves and provides assignments for some of the inner valence features, as noted.

For oxazolidinone, Figure 7, theory predicts that the first band consists of three ionic states A (23a), B (22a) and C (21a), and this explains its unusual shape, in which two peaks are clearly distinguished (A and C), while the remaining peak (B) appears to be a shoulder, but might have been due to vibrational structure. In a previous study with He I radiation,[23] this region of the spectrum appeared to consist of only two peaks. The fitted widths of these three states vary from 230 to 360 meV, or 113 to 300 meV after subtracting the resolution in quadrature. These very narrow line widths, of the order of a single vibrational quantum, suggest that the Franck-Condon envelopes of these states contain very few vibrational states. From this we may conclude that ground and excited state potential energy curves and equilibrium bond distances are very similar, and that these orbitals have mostly non-bonding character.

The next state D is assigned to ionization of orbital 20a. Features E and F are asymmetric and can be fitted with only two peaks. However theory indicates that four ionic states contribute to this structure: ionization of orbitals 19a to 16a. The intensity of features E and F, much higher than the preceding or following peaks, is qualitatively consistent with the assignment to ionization of more than two orbitals. Similarly peak G is assigned to a single state, while peak H is due to two states. The higher binding energy peaks are not calculated by the OVGF method, which considers only the outer valence, so they are assigned as shown in Table 3 on the basis of the SAOP calculations. In this energy range, there are many two-hole, single-particle states, so the calculations provide only the main character of the stronger spectral features.

The average difference between the calculated OVGF ionization potentials and the experimental values is 0.2 eV. For the SAOP calculations, the difference is larger, 0.40 eV if we include only the peaks calculated using OVGF, or 0.59 eV if all calculated orbitals are included. Also, the discrepancies are large for the HOMO and HOMO-1 (SAOP orbitals), where the ionization potentials are overestimated, but the agreement improves for intermediate orbitals. For inner valence orbitals, the discrepancy is again large, but the sign of the difference changes and the SAOP calculations underestimate the energies.

The orbital character of the valence states can be determined by mapping the ground state charge density of the orbitals, Figure 8. The HOMO of both conformers of CS consists of an extended molecular orbital with π character involving the C=O bond, but extending over all heavy atoms of the heterocycle. The HOMO of OX2 is also based on atomic orbitals of $N_{(3)}$ pπ, $O_{(6)}$ pπ and $C_{(4)}$ pπ character, but is more heavily localised on the $N_{(3)}$-$C_{(4)}$ bond than the other heavy atoms. The HOMO-1 orbitals of the two conformers of CS show much more significant differences than the pairs of orbitals above or below it in energy. In conformer I the orbital is delocalized over most of the molecule, but in conformer II, it is localized on the primary amino group with strong $N_{(7)}$ pπ character. The HOMO-1 of OX2 does not appear to be similar in character to the HOMO-1 orbitals of the CS conformers.

The HOMO-2 of CS-**I** is largely localised on $O_{(6)}$-$C_{(4)}$-$N_{(7)}$, while that of conformer **II** is delocalised over the molecule. The corresponding orbital of OX2 is composed of p type orbitals localized on $O_{(1)}$-$C_{(5)}$-$C_{(3)}$. Considering CS as an isomerised derivative of OX2, the orbitals appear to undergo two changes. Firstly there is a change in the energy ordering, as the HOMO-2 of OX2 most closely resembles conformer **II** of CS. Similarly the HOMO-1 of OX2 shows some resemblance to the HOMO-3 of conformer **I**. Secondly there is a stronger effect of mixing and rearrangement of the orbital character so that the orbitals of CS become distinct from those of OX2.

As mentioned above, we have evidence for resonance phenomena from the geometrical structure and we therefore expect that some molecular orbitals will have π character and be localised on the atoms displaying resonance. We examined the character of the valence orbitals and we find the orbitals shown in Figure 9, orbital 18a (HOMO-9) of CS-**I** and orbital 18a (HOMO-5) of OX2, have the expected charge distribution. In CS, the molecular orbital is made up of atomic $p_z$ contributions from $O_{(1)}$, $N_{(2)}$, the carbonyl group and $N_{(7)}$, while in OX2, the corresponding orbitals of the carbonyl group and $N_{(3)}$ are mixed, with a contribution from the lone pair of $C_{(5)}$. Thus the valence molecular orbital charge distribution, the bonding geometry and the core level spectra are consistent with the existence of resonance effects in both molecules.

Interestingly, the effect of conformerism, which involves simple twisting of covalent bonds and formation or breaking of hydrogen bonds, has quite a strong effect on the character of the molecular orbitals. Presumably this is primarily because the extent of spatial overlaps of available atomic orbitals changes.

## 4. CONCLUSIONS

In the present study, the core and valence soft X-ray photoelectron spectra of cycloserine and 2-oxazolidinone have been determined and assigned applying high quality quantum mechanical calculations. The character of the frontier valence orbitals has been assigned, and the core level spectra related to the electronic structure. Although theory predicted that two conformers of cycloserine were populated, we were unable to find unambiguous experimental evidence of this. On the other hand, theoretical analysis predicted resonance effects in the valence bonding, and identified the molecular orbitals involved. Core level spectra and published bond lengths support this interpretation, leading to a consistent picture of the electronic structure.

## ACKNOWLEDGMENTS

The authors wish to thank Professor D. P. Chong for his assistance in the ΔDFT calculations. MA acknowledges the Swinburne University Postgraduate Research Award (SUPRA). FW thanks the National Computational Infrastructure (NCI) at the Australian National University under the Merit Allocation Scheme (MAS), the Victorian Partnership for Advanced Computing (VPAC) and Swinburne University supercomputing (Green/gSTAR). We thank our colleagues at Elettra-Sincrotrone Trieste for providing high quality synchrotron light.

**Supporting Information Available**
The CS potential energy surface scan together with the relative energy and populations of the most stable CS conformers.

# REFERENCES

[1] M. Grzemska, T. Mihaescu, L. Clancy, and L. Casali, Eur. Respir. J. **14**, 978 (1999).
[2] K. Mdluli and M. Spigelman, Curr. Opin. Pharmacol. **6**, 459 (2006).
[3] S. Tsiodras, H.S. Gold, G. Sakoulas, G.M. Eliopoulos, C. Wennersten, L. Venkataraman, R.C. Moellering, and M.J. Ferraro, Lancet **358**, 207 (2001).
[4] L.S. Dennis, H. Daniel, L. Harry, L.H. John, H.B. Donald, and H. Barry, Clin. Infect. Dis. **34**, 1481 (2002).
[5] Z. Li, R.J. Willke, L.A. Pinto, B.E. Rittenhouse, M.J. Rybak, A.M. Pleil, C.W. Crouch, B. Hafkin, and H.A. Glick, Pharmacotherapy **21**, 263 (2001).
[6] B.B. Lohray, S. Baskaran, B. Srinivasa Rao, B. Yadi Reddy, and I. Nageswara Rao, Tetrahedron Lett. **40**, 4855 (1999).
[7] D.K. Hutchinson, Curr. Top. Med. Chem. **3**, 1021 (2003).
[8] A.P.W. Arachchilage, F. Wang, V. Feyer, O. Plekan, and K.C. Prince, J. Chem. Phys. **133**, 174319 (2010).
[9] A.P.W. Arachchilage, F. Wang, V. Feyer, O. Plekan, and K.C. Prince, J. Chem. Phys. **136**, 124301 (2012).
[10] M. Ahmed, A. Ganesan, F. Wang, V. Feyer, O. Plekan, and K.C. Prince, J. Phys. Chem. A **116**, 8653 (2012).
[11] J.W. Turley and R. Pepinsky, Acta Cryst. **10**, 480 (1957).
[12] H.-H. Lee, H. Yamaguchi, H. Senda, S. Maeda, A. Kuwae, and K. Hanai, Spectros. Lett. **30**, 685 (1997).
[13] H.-H. Lee, N. Takeuchi, H. Senda, A. Kuwae, and K. Hanai, Spectros. Lett. **31**, 1217 (1998).
[14] J. Yosa, M. Blanco, O. Acevedo, and L.R. Lareo, Eur. J. Med. Chem. **44**, 2960 (2009).
[15] S.H. Gerson, S.D. Worley, N. Bodor, J.J. Kaminski, and T.W. Flechtner, J. Electron. Spectrosc. Relat. Phenom. **13**, 421 (1978).
[16] S. Okumoto and S. Yamabe, J. Comput. Chem. **22**, 316 (2001).
[17] V.A. Pankratov, Ts.M.Frenkel, and A.M. Fainleib, Russ. Chem. Rev. **52**, 576 (1983).
[18] D. Kaur and R. Sharma, Struct. Chem. **23**, 905 (2012).
[19] I. Konstantinov and L. Broadbelt, Top. Catal. **53**, 1031 (2010).
[20] A.K. Sharma and R.B. Sunoj, Angew. Chem. Int. Ed. **49**, 6373 (2010).
[21] J. Wouters, F. Ooms, and F. Durant, Acta Crystallogr. C **53**, 895 (1997).
[22] H.T. Flakus and A. Michta, Vib. Spectrosc. **33**, 177 (2003).
[23] M.V. Andreocci, F.A. Devillanova, C. Furlani, G. Mattogno, G. Verani, and R. Zanoni, J. Mol. Struct. **69**, 151 (1980).
[24] M.V. Andreocci, M. Bossa, F.A. Devillanova, C. Furlani, G. Mattogno, G. Verani, and R. Zanoni, J. Mol. Struct. **71**, 227 (1981).
[25] O. Plekan, V. Feyer, R. Richter, M. Coreno, M. de Simone, K.C. Prince, and V. Carravetta, Chem. Phys. Lett. **442**, 429 (2007).
[26] O. Plekan, V. Feyer, R. Richter, M. Coreno, M. de Simone, K.C. Prince, and V. Carravetta, J. Phys. Chem. A **111**, 10998 (2007).
[27] O. Plekan, V. Feyer, R. Richter, M. Coreno, M. de Simone, K.C. Prince, and V. Carravetta, J. Electron. Spectrosc. Relat. Phenom. **155**, 47 (2007).
[28] V. Feyer, O. Plekan, R. Richter, M. Coreno, M. de Simone, K.C. Prince, A.B. Trofimov, I.L. Zaytseva, and J. Schirmer, J. Phys. Chem. A **114**, 10270 (2010).
[29] F. Weinhold and C.R. Landis, Chem. Educ. Res. Pract. **2**, 91 (2001).
[30] Gaussian 09, Revision A.02, M. J. Frisch, G. W. Trucks, H. B. Schlegel, G. E. Scuseria, M. A. Robb, J. R. Cheeseman, G. Scalmani, V. Barone, B. Mennucci, G. A. Petersson, H. Nakatsuji, M. Caricato, X. Li, H. P. Hratchian, A. F. Izmaylov, J. Bloino, G. Zheng, J. L. Sonnenberg, M. Hada, M. Ehara, K. Toyota, R. Fukuda, J. Hasegawa, M. Ishida, T. Nakajima, Y. Honda, O. Kitao, H. Nakai, T. Vreven, J. A. Montgomery, Jr., J. E. Peralta, F. Ogliaro, M. Bearpark, J. J. Heyd, E. Brothers, K. N. Kudin, V. N. Staroverov, R. Kobayashi, J. Normand, K. Raghavachari, A. Rendell, J. C. Burant, S. S. Iyengar, J. Tomasi, M. Cossi, N. Rega, J. M. Millam, M. Klene, J. E. Knox, J. B. Cross, V. Bakken, C. Adamo, J. Jaramillo, R. Gomperts, R. E. Stratmann, O. Yazyev, A. J. Austin, R. Cammi, C. Pomelli, J. W. Ochterski, R. L. Martin, K. Morokuma, V. G. Zakrzewski, G. A. Voth, P. Salvador, J. J. Dannenberg, S. Dapprich, A. D. Daniels, Ö. Farkas, J. B. Foresman, J. V. Ortiz, J. Cioslowski, and D. J. Fox, Gaussian, Inc., Wallingford CT, **2009**.
[31] C.T. Falzon and F. Wang, J. Chem. Phys. **123**, 214307 (2005).
[32] C.T. Falzon, F. Wang, and W. Pang, J. Phys. Chem. B **110**, 9713 (2006).
[33] A. Ganesan and F. Wang, J. Chem. Phys. **131**, 044321 (2009).

[34] J. Schirmer, L.S. Cederbaum, and O. Walter, Phys. Rev. A **28**, 1237 (1983).
[35] D. Danovich, WIREs Comput. Mol. Sci. **1**, 377 (2011).
[36] A.P.W. Arachchilage, F. Wang, V. Feyer, O. Plekan, and K.C. Prince, J. Chem. Phys. **133**, 174319 (2010).
[37] A. Ganesan, F. Wang, M. Brunger, and K. Prince, J. Synchrotron Radiat. **18**, 733 (2011).
[38] L. Selvam, V. Vasilyev, and F. Wang, J. Phys. Chem. B **113**, 11496 (2009).
[39] F. Chen and F. Wang, Molecules **14**, 2656 (2009).
[40] G. te Velde, F.M. Bickelhaupt, E.J. Baerends, C. Fonseca Guerra, S.J.A. van Gisbergen, J.G. Snijders, and T. Ziegler, J. Comput. Chem. **22**, 931 (2001).
[41] C. Fonseca Guerra, J.G. Snijders, G. te Velde, and E.J. Baerends, Theor. Chem. Acc. **99**, 391 (1998).
[42] ADF2010, SCM, Theoretical Chemistry, Vrije University, Amsterdam, The Netherlands.
[43] Y. Takahata, C.E. Wulfman, and D.P. Chong, J. Mol. Struct. (THEOCHEM) **863**, 33 (2008).
[44] R. van Leeuwen and E.J. Baerends, Phys. Rev. A **49**, 2421 (1994).
[45] P.R.T. Schipper, O.V. Gritsenko, S.J.A.v. Gisbergen, and E.J. Baerends, J. Chem. Phys. **112**, 1344 (2000).
[46] D.P. Chong, E. Van Lenthe, S. Van Gisbergen, and E.J. Baerends, J. Comput. Chem. **25**, 1030 (2004).
[47] T. Koopmans, Physica 1 **1**, 104 (1933).
[48] GaussView, Version 5, R. Dennington, T. Keith and J. Millam, Semichem Inc., Shawnee Mission KS, 2009.
[49] G. Schaftenaar and J.H. Noordik, J. Comput.-Aided Mol. Des. **14**, 123 (2000).
[50] E. Westhof and M. Sundaralingam, J. Am. Chem. Soc. **105**, 970 (1983).
[51] C. Altona and M. Sundaralingam, J. Am. Chem. Soc. **94**, 8205 (1972).
[52] P. Bolognesi, G. Mattioli, P. O'Keeffe, V. Feyer, O. Plekan, Y. Ovcharenko, K.C. Prince, M. Coreno, A. Amore Bonapasta, and L. Avaldi, J. Phys. Chem. A **113**, 13593 (2009).

**Table 1:** Comparison of selected geometric parameters of CS (three conformers) and OX2 with available theoretical and experimental crystal structure data.

| Cycloserine | | | | | 2-oxazolidinone | | | |
|---|---|---|---|---|---|---|---|---|
| **Parameters** | **Theory** | | | **Expt.**[a] | **Parameters** | **Theory** | **Kaur et al.**[b] | **Expt.**[c] |
| | CS-I | CS-II | CS-III | | | | | |
| $R_5$ (Å) | 7.322 | 7.308 | 7.339 | 7.218 | $R_5$ (Å) | 7.184 | 7.208 | 7.073 |
| $O_{(1)}$-$N_{(2)}$ (Å) | 1.426 | 1.422 | 1.427 | 1.458 | $O_{(1)}$-$C_{(2)}$ (Å) | 1.371 | 1.377 | 1.356 |
| $N_{(2)}$-$C_{(3)}$ (Å) | 1.378 | 1.373 | 1.383 | 1.299 | $C_{(2)}$-$N_{(3)}$ (Å) | 1.381 | 1.387 | 1.301 |
| $C_{(3)}$-$C_{(4)}$ (Å) | 1.546 | 1.539 | 1.549 | 1.519 | $N_{(3)}$-$C_{(4)}$ (Å) | 1.455 | 1.470 | 1.466 |
| $C_{(4)}$-$C_{(5)}$ (Å) | 1.526 | 1.531 | 1.538 | 1.502 | $C_{(4)}$-$C_{(5)}$ (Å) | 1.538 | 1.528 | 1.497 |
| $C_{(5)}$-$O_{(1)}$ (Å) | 1.445 | 1.442 | 1.442 | 1.440 | $C_{(5)}$-$O_{(1)}$ (Å) | 1.440 | 1.446 | 1.453 |
| $O_{(1)}$-$N_{(2)}$-$C_{(3)}$ (°) | 111.24 | 111.54 | 111.26 | 107.7 | $O_{(1)}$-$C_{(2)}$-$N_{(3)}$ (°) | 108.12 | 108.23 | 110.2 |
| $N_{(2)}$-$C_{(3)}$-$C_{(4)}$ (°) | 106.13 | 106.49 | 105.80 | 112.2 | $C_{(2)}$-$N_{(3)}$-$C_{(4)}$ (°) | 111.09 | 103.95 | 113.3 |
| $C_{(3)}$-$C_{(4)}$-$C_{(5)}$ (°) | 101.12 | 101.53 | 100.62 | 100.3 | $N_{(3)}$-$C_{(4)}$-$C_{(5)}$ (°) | 99.36 | 98.85 | 100.3 |
| $C_{(4)}$-$C_{(5)}$-$O_{(1)}$ (°) | 104.85 | 105.88 | 104.53 | 103.9 | $C_{(4)}$-$C_{(5)}$-$O_{(1)}$ (°) | 104.49 | 103.95 | 106.0 |
| $C_{(5)}$-$O_{(1)}$-$N_{(2)}$ (°) | 102.61 | 103.16 | 102.00 | 107.0 | $C_{(5)}$-$O_{(1)}$-$C_{(2)}$ (°) | 109.43 | 108.44 | 108.6 |
| $\nu_{max}$ | 37.96 | 34.49 | 40.16 | 29.33 | $\nu_{max}$ | 26.49 | | |
| $<R^2>$ (a.u.) | 648.289 | 649.945 | 651.573 | | $<R^2>$ (a.u.) | 476.378 | | |
| $\mu$ (D) | 1.8548 | 2.789 | 3.853 | | $\mu$ (D) | 5.519 | | |
| Total Energy (Hartree) | -377.934 | -377.932 | -377.929 | | Total Energy (Hartree) | -322.638 | | |
| $ZPE^d$ (kJ·mol$^{-1}$) | 270.153 | 269.726 | 268.998 | | ZPE (kJ·mol$^{-1}$) | 229.124 | | |
| $TE^e$ +ZPE (Hartree) | -377.831 | -377.830 | -377.827 | | TE+ZPE (Hartree) | -322.551 | | |
| ΔTE (kJ·mol$^{-1}$) | 0.00 | 5.006 | 13.254 | | ΔTE (kJ·mol$^{-1}$) | | | |
| ΔTE +ZPE (kJ·mol$^{-1}$) | 0.00 | 4.582 | 12.010 | | ΔTE +ZPE (kJ·mol$^{-1}$) | | | |

[a] X-ray crystal structure, see [13].
[b] B3LYP/6-31+G*, see [18].
[c] X-ray Crystal structure, see [17].
[d] ZPE: Zero Point Energy.
[e] TE: Total energy.

**Table 2**: Comparison between measured and simulated core electrons ionization potentials of CS (two conformers) and OX2 in eV.

| Atomic site | CS | | | | | Atomic site | OX2 | | |
|---|---|---|---|---|---|---|---|---|---|
| | Theory | | | | Expt. | | Theory | | Expt. |
| | CS-**I** | | CS-**II** | | | | | | |
| | Vertical[a] | $\Delta E_{ks}$[b] | Vertical[a] | $\Delta E_{ks}$[b] | | | Vertical[a] | $\Delta E_{ks}$[b] | |
| $C_{(3)}$ | 292.76 | 293.50 | 292.83 | 293.67 | 293.9 | $C_{(2)}$ | 293.61 | 294.80 | 295.20 |
| $C_{(4)}$ | 291.41 | 292.11 | 291.38 | 292.08 | 291.9 | $C_{(4)}$ | 291.53 | 292.58 | 292.45 |
| $C_{(5)}$ | 291.74 | 292.64 | 291.68 | 292.58 | 292.3 | $C_{(5)}$ | 291.93 | 292.91 | 292.75 |
| $N_{(2)}$ | 405.73 | 407.10 | 405.73 | 407.14 | 407.2 | $N_{(3)}$ | 404.34 | 406.20 | 406.2 |
| $N_{(7)}$ | 403.37 | 405.48 | 403.22 | 405.41 | 405.6 | | | | |
| $O_{(1)}$ | 536.63 | 539.54 | 536.53 | 539.47 | 539.85 | $O_{(1)}$ | 536.11 | 539.15 | 539.46 |
| $O_{(6)}$ | 534.58 | 537.23 | 534.62 | 537.36 | 537.75 | $O_{(6)}$ | 534.27 | 537.26 | 537.55 |

[a]Vertical IPs using the LB94/et-PVQZ model applying the "meta-Koopmans" theorem.
[b]$\Delta E_{ks}$ (PW86-PW91)/et-pVQZ.

**Table 3:** Comparison between measured and simulated valence electron ionization potentials of CS (two conformers) and OX2 in eV.

| | Cycloserine | | | | | 2-oxazolidinone | | | | | |
| | CS-I | | CS-II | | | | | | | | |
| Orbital | SAOP[a] | OVGF[b] | SAOP[a] | OVGF[b] | Expt. | Orbital | SAOP[a] | OVGF[b,] | Expt.[c] | Expt.[d] | Expt. |
|---|---|---|---|---|---|---|---|---|---|---|---|
| 28a(LUMO) | 5.72 | | 5.76 | | | 24a(LUMO) | 4.60 | | | | |
| 27a (HOMO) | 10.10 | 9.14 | 10.31 | 9.23 | 9.22 (A) | 23a (HOMO) | 11.15 | 10.08 | 10.06 | 10.21 | 10.10 (A) |
| 26a (HOMO-1) | 10.70 | 9.86 | 10.62 | 10.04 | 9.88 (B) | 22a (HOMO-1) | 11.46 | 10.95 | 10.98 | 10.71 | 10.50 (B) |
| 25a (HOMO-2) | 11.54 | 11.19 | 11.14 | 10.52 | 10.82 (C) | 21a (HOMO-2) | 11.94 | 11.21 | | 11.07 | 11.04 (C) |
| 24a | 12.82 | 12.41 | 12.84 | 12.35 | 11.94 (D) | 20a | 13.52 | 13.15 | 13.0 | 12.82 | 13.03 (D) |
| 23a | 13.43 | 13.09 | 13.84 | 13.53 | 13.28 (E) | 19a | 14.67 | 14.55 ⎫ | | | 14.21 (E) |
| 22a | 14.53 | 14.49 | 14.27 | 14.21 | 13.96 (F) | 18a | 14.86 | 14.80 ⎪ | | | |
| 21a | 14.57 | 14.60 | 14.67 | 14.56 | 14.53 (G) | 17a | 15.02 | 15.01 ⎬ | | | 14.85 (F) |
| 20a | 15.13 | 15.02 | 15.13 | 15.01 | - | 16a | 15.18 | 15.21 ⎭ | | | |
| 19a | 15.55 | 15.46 | 15.24 | 15.01 | - | 15a | 16.43 | 16.48 | | | 16.36 (G) |
| 18a | 15.72 | 15.79 | 15.69 | 15.84 | 15.74 (H) | 14a | 17.51 | 17.79 ⎤ | | | 17.65 (H) |
| 17a | 16.34 | 16.58 | 16.74 | 17.06 | - | 13a | 17.68 | 17.92 ⎦ | | | |
| 16a | 17.40 | 17.65 | 17.13 | 17.37 | - | 12a | 19.77 | | | | 19.92 (I) |
| 15a | 17.61 | 17.81 | 17.67 | 17.77 | 17.98 (I) | 11a | 20.82 | | | | 21.5 (J) |
| 14a | 19.68 | | 19.56 | | 19.61 (J) | 10a | 23.64 | | | | 24.5 (K) |
| 13a | 20.34 | | 20.41 | | 20.76 (K) | 9a | 27.95 | | | | 29.1 (L) |
| 12a | 23.44 | | 23.45 | | 24.18 (L) | 8a | 30.34 | | | | 31.5 (M) |
| 11a | 26.13 | | 26.16 | | - | 7a | 32.61 | | | | 34 (N) |
| 10a | 27.07 | | 27.00 | | 27.6 (M) | | | | | | |
| 9a | 30.72 | | 30.80 | | - | | | | | | |
| 8a | 32.52 | | 32.51 | | 33.2 (N) | | | | | | |
| | | | | | | | | | | | |
| HOMO-LUMO gap | 4.39 | | 4.55 | | | HOMO-LUMO gap | 6.55 | | | | |

[a] SAOP/et-pVQZ.

[b] OVGF/6-311++G**.

[c] Andreocci et al. [23]

[d] Gerson et al.[15].

**Figure captions.**

**Figure 1**: Schematic structure of (a) cycloserine and (b) 2-oxazolidinone, showing numbering of atoms.

**Figure 2:** The 3-dimensional ball and stick models of the ground state structures with potential intramolecular H-bonds and with NBO charges in parentheses of (a) cycloserine (CS-**I**) (b) cycloserine (CS-**II**) (c) cycloserine (CS-**III**) (d) 2-oxazolidinone. Colors represent: red, oxygen; blue, nitrogen; dark grey, carbon; light grey: hydrogen.

**Figure 3:** Experimental and simulated C1s core ionization spectra of OX2 (lower panel, theoretical FWHM=0.3) and CS-**I** (upper panel, theoretical FWHM=0.5), using the $\Delta E_{ks}$ (PW86-PW91)/et-pVQZ level of theory.

**Figure 4:** Experimental and simulated N1s core ionization spectra of OX2 (lower panel) and CS-**I** (upper panel) using the $\Delta E_{ks}$ (PW86-PW91)/et-pVQZ level of theory and FWHM = 0.5 eV.

**Figure 5:** Experimental and simulated O1s core ionization spectra of OX2 (lower panel, simulated spectrum is shifted by +0.3 eV to match experiment) and CS-**I** (upper panel, simulated spectrum is shifted by +0.5 eV to match experiment), using the $\Delta E_{ks}$ (PW86-PW91)/et-pVQZ level of theory and FWHM = 0.9 eV.

**Figure 6:** Experimental and OVGF/6-311++G** simulated valence band photoelectron spectra of cycloserine. Inset: inner valence band. Photon energy: 100 eV. The simulated spectra of conformers are weighted by their calculated Boltzmann factors at the experimental temperature. The simulated spectrum is broadened by FWHM=0.25 eV.

**Figure 7:** Experimental and OVGF/6-311++G** simulated valence band photoelectron spectra of oxazolidinone. Inset: inner valence band. Photon energy: 100 eV. The simulated spectrum is broadened by FWHM=0.25 eV.

**Figure 8:** Electronic distribution of the frontier orbitals of CS and OX2, SAOP/et-pVQZ.

**Figure 9**: Left: orbital 18a (HOMO-9) of CS-**I**. Right: orbital 18a (HOMO-5) in OX2, SAOP/et-pVQZ.

**Figure 1**: Schematic structure of (a) cycloserine and (b) 2-oxazolidinone, showing numbering of atoms.

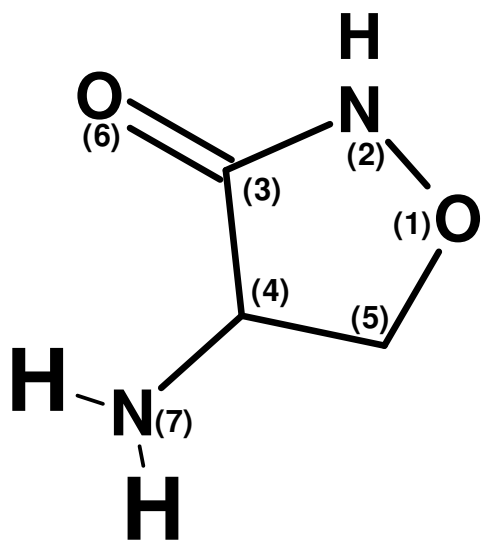
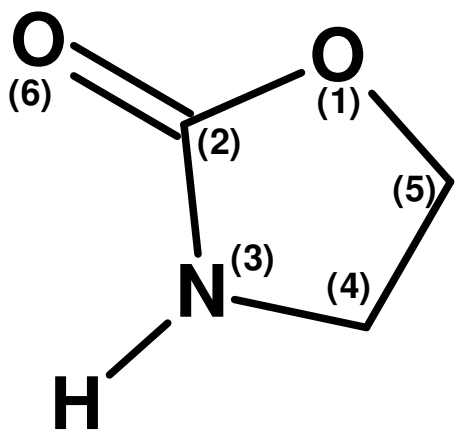

(a)Cycloserine.　　　　　　　　(b)2-oxazolidinone.

**Figure 2:** The 3-dimensional ball and stick models of the ground state structures with potential intramolecular H-bonds and with NBO charges in parentheses of (a) cycloserine (CS-**I**) (b) cycloserine (CS-**II**) (c) cycloserine (CS-**III**) (d) 2-oxazolidinone. Colors represent: red, oxygen; blue, nitrogen; dark grey, carbon; light grey: hydrogen.

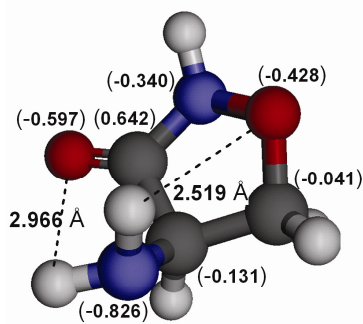
(a) Cycloserine (CS-**I**)

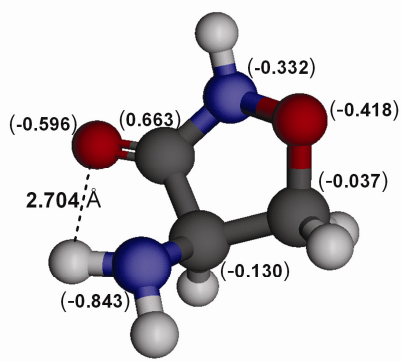
(b) Cycloserine (CS-**II**)

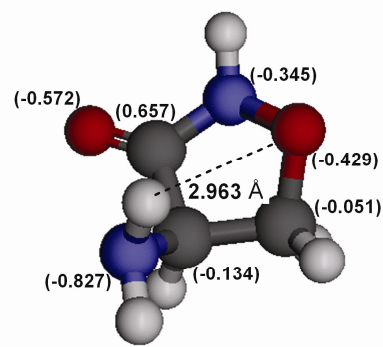
(c) Cycloserine (CS-**III**)

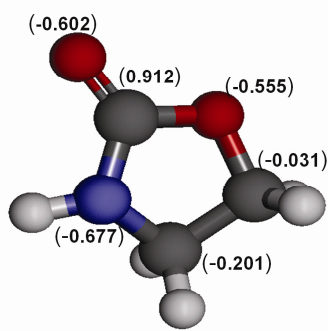
(d) 2-oxazolidinone

**Figure 3:** Experimental and simulated C1s core ionization spectra of OX2 (lower panel, theoretical FWHM=0.3) and CS-**I** (upper panel, theoretical FWHM=0.5), using the $\Delta E_{ks}$ (PW86-PW91)/et-pVQZ level of theory.

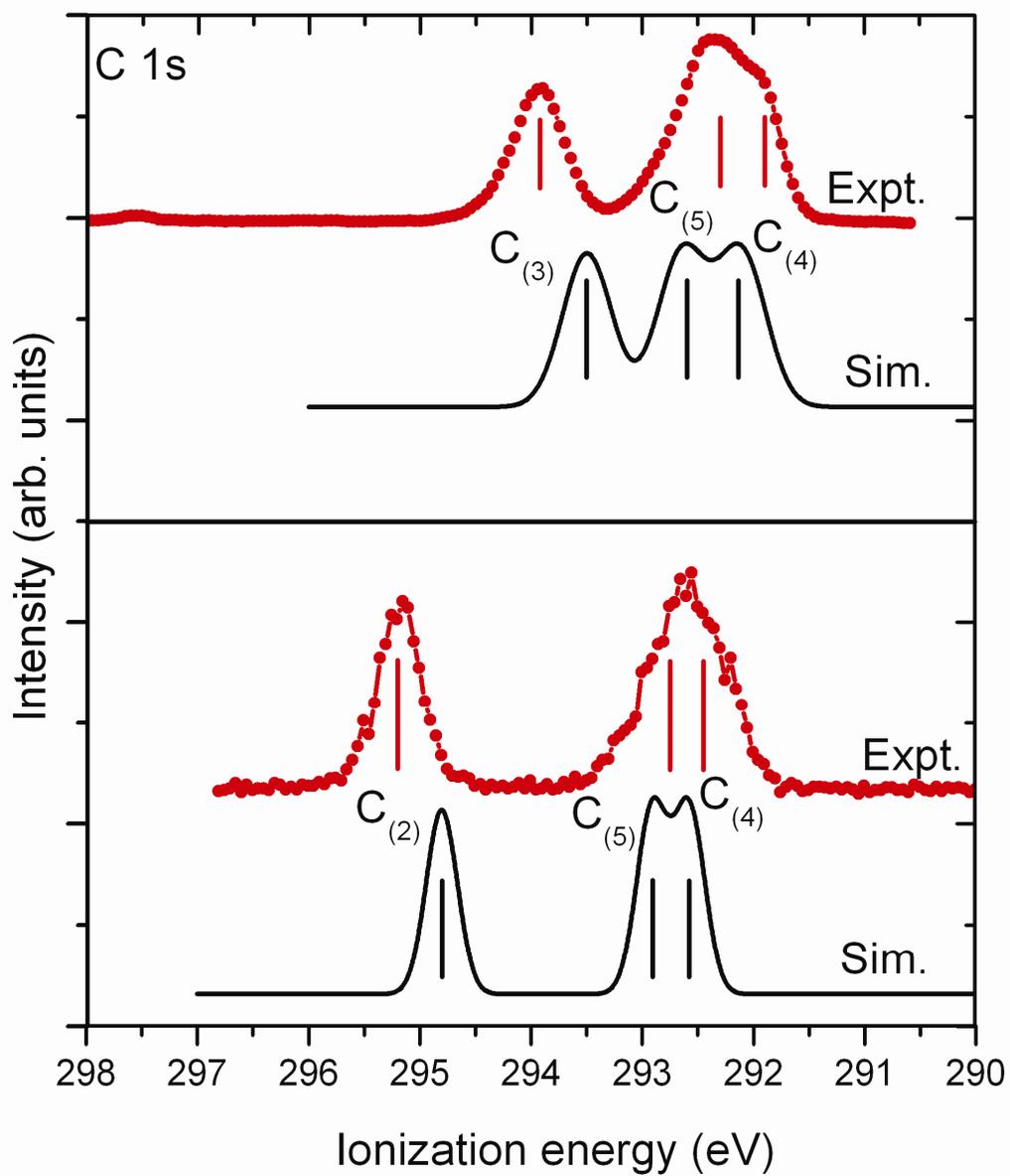

**Figure 4:** Experimental and simulated N1s core ionization spectra of OX2 (lower panel) and CS-I (upper panel) using the $\Delta E_{ks}$ (PW86-PW91)/et-pVQZ level of theory and FWHM = 0.5 eV.

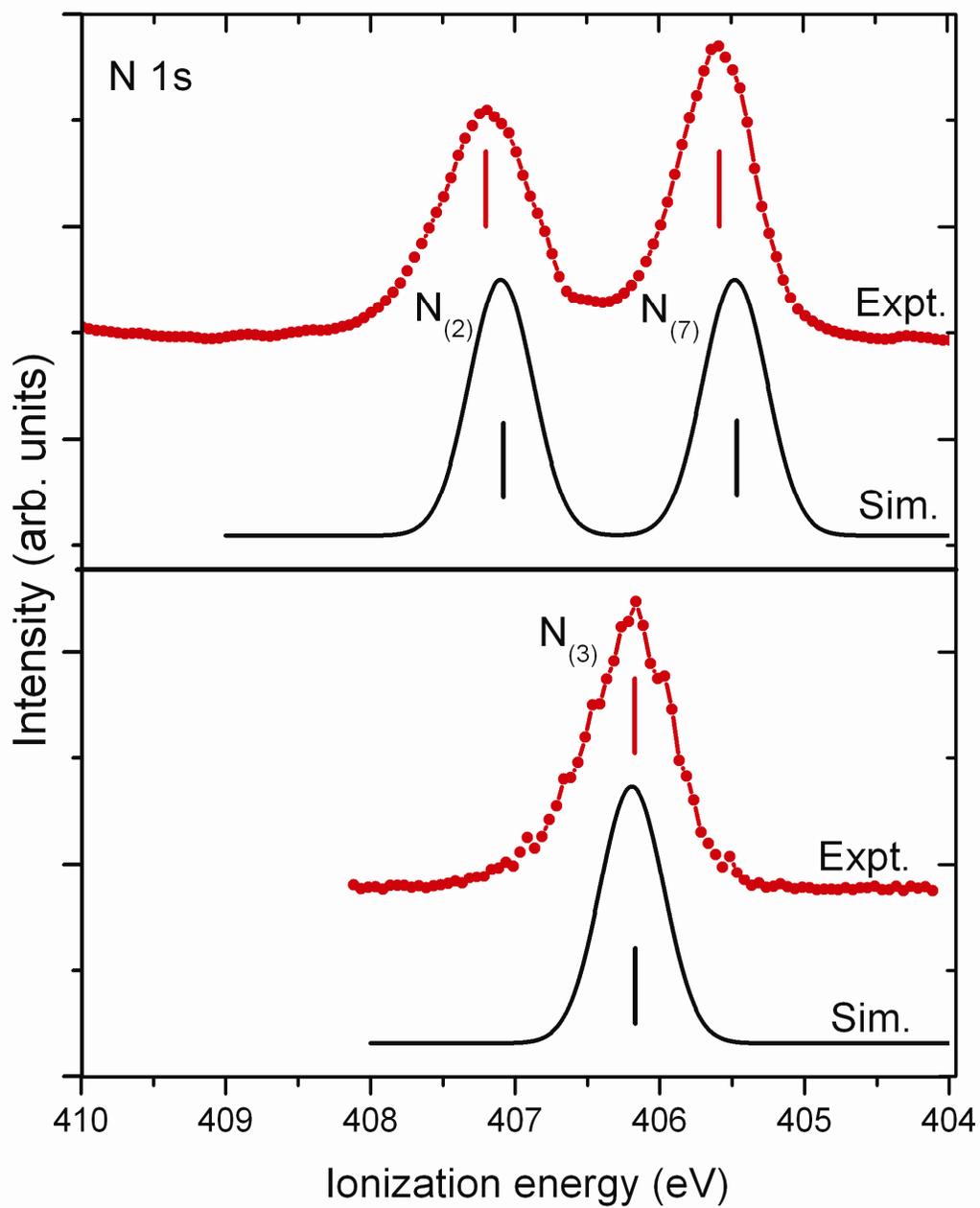

**Figure 5:** Experimental and simulated O1s core ionization spectra of OX2 (lower panel, simulated spectrum is shifted by +0.3 eV to match experiment) and CS-**I** (upper panel, simulated spectrum is shifted by +0.5 eV to match experiment), using the $\Delta E_{ks}$ (PW86-PW91)/et-pVQZ level of theory and FWHM = 0.9 eV.

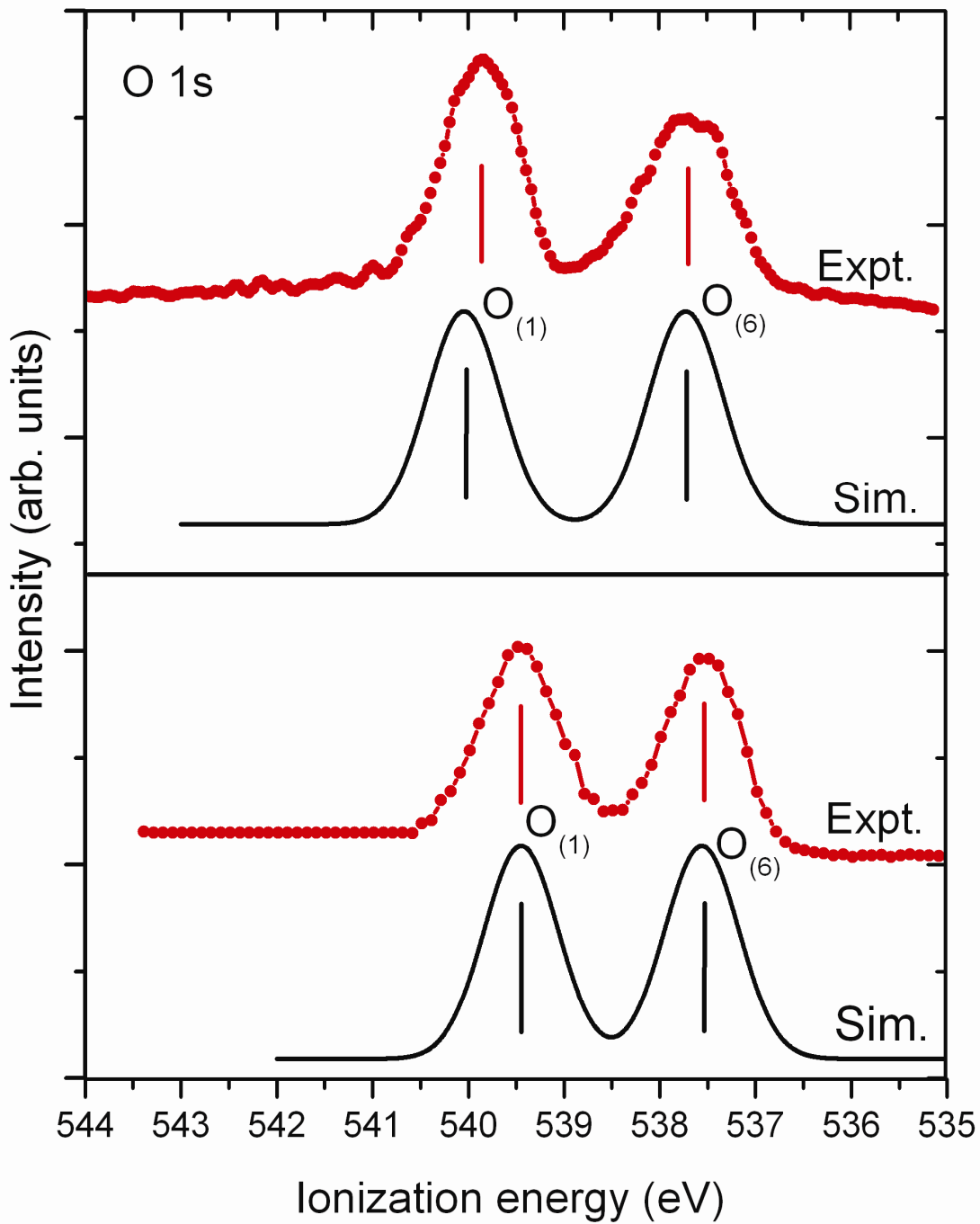

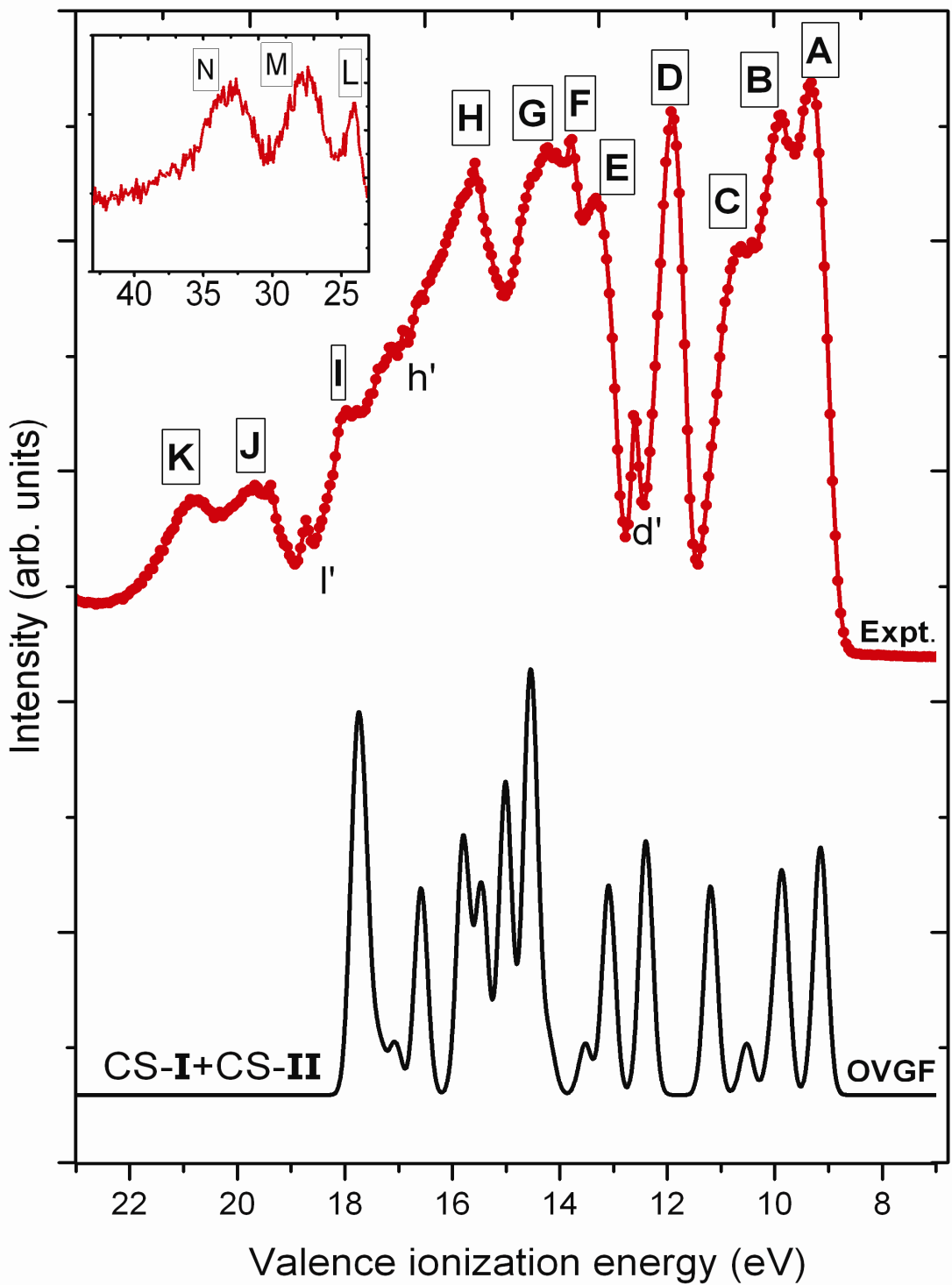

**Figure 6:** Experimental and OVGF/6-311++G** simulated valence band photoelectron spectra of cycloserine. Inset: inner valence band. Photon energy: 100 eV. The simulated spectra of conformers are weighted by their calculated Boltzmann factors at the experimental temperature. The simulated spectrum is broadened by FWHM=0.25 eV.

**Figure 7:** Experimental and OVGF/6-311++G** simulated valence band photoelectron spectra of oxazolidinone. Inset: inner valence band. Photon energy: 100 eV. The simulated spectrum is broadened by FWHM=0.25 eV.

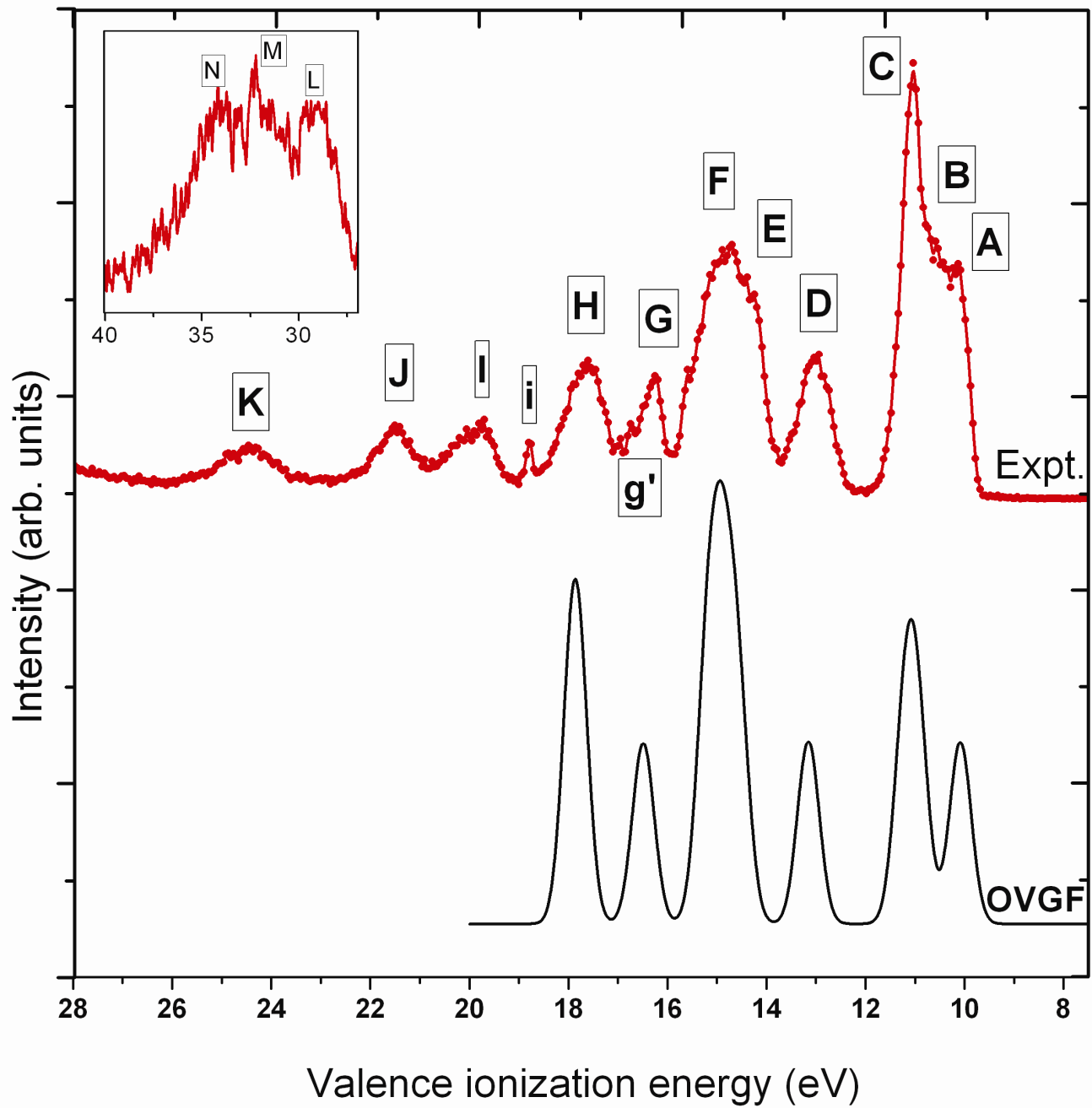

**Figure 8:** Electronic distribution of the frontier orbitals of CS and OX2, SAOP/et-pVQZ.

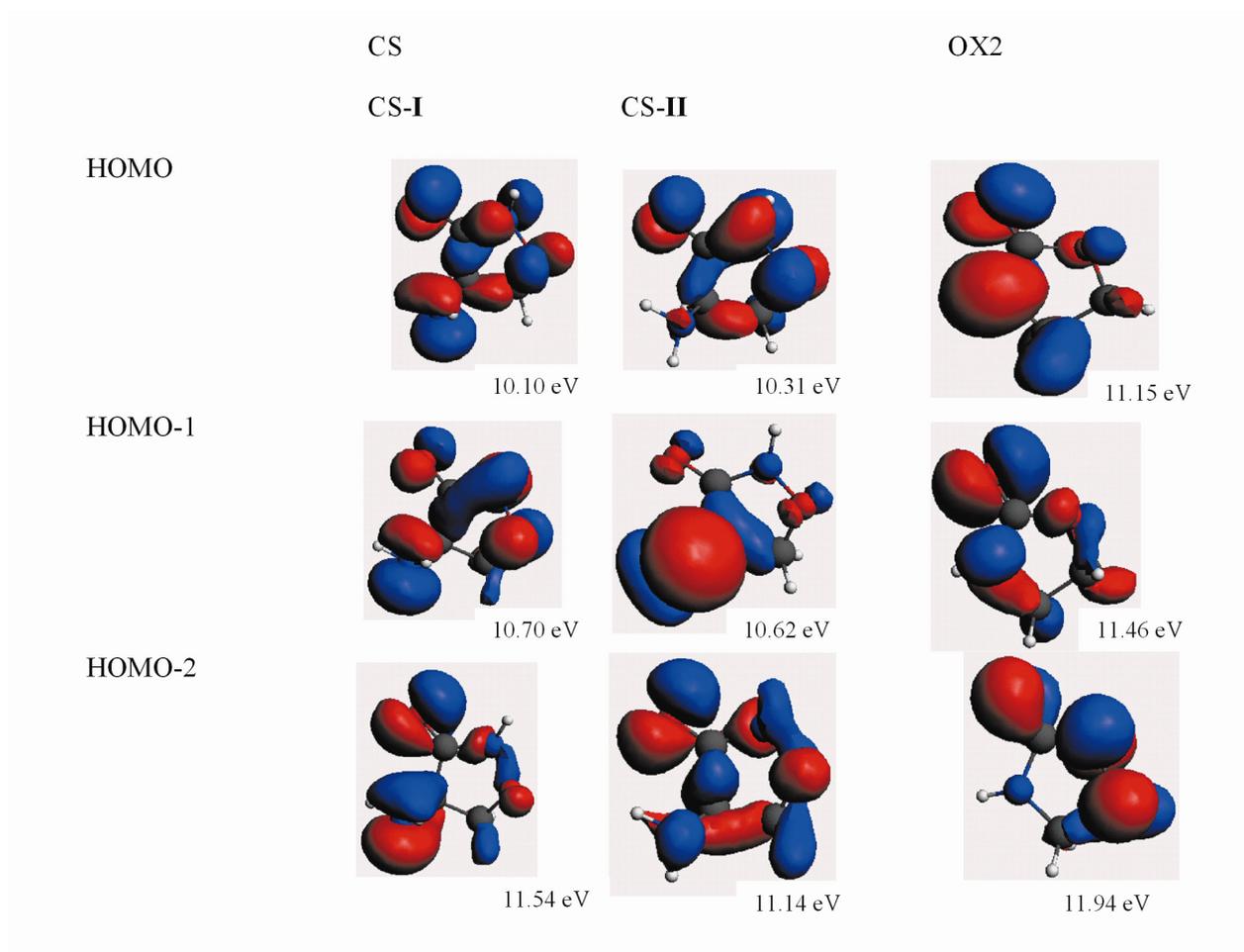

**Figure 9**: Left: orbital 18a (HOMO-9) of CS-**I**. Right: orbital 18a (HOMO-5) in OX2, SAOP/et-pVQZ.

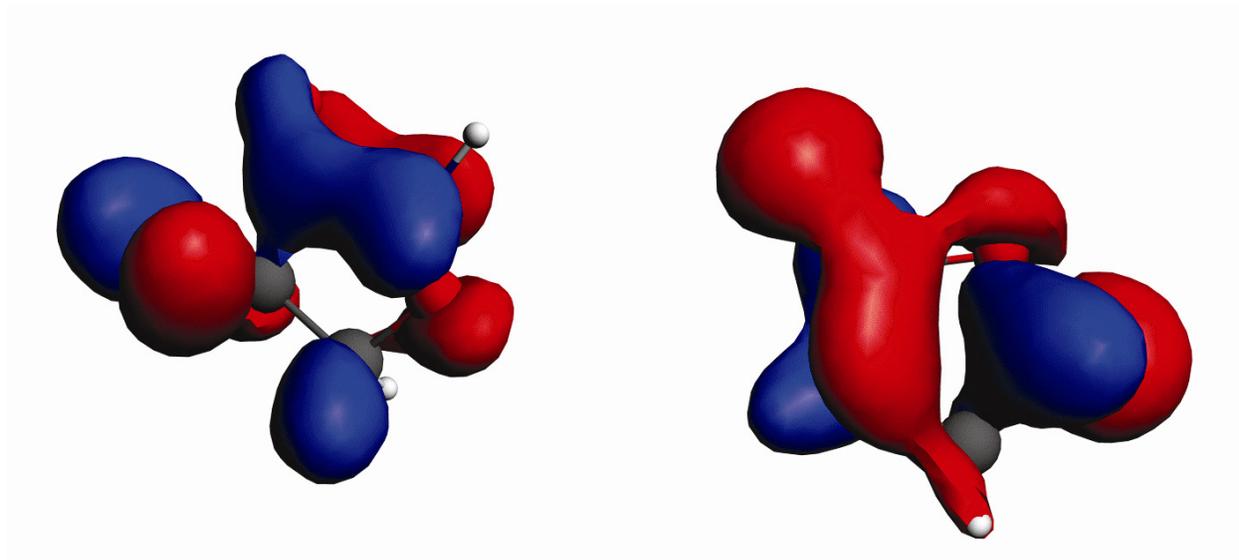